\documentclass[aps,prl,twocolumn]{revtex4}

\begin{document}

\title{Existence of Density Functionals for Excited States and Resonances} 

\author{B.G. Giraud}
  \email{giraud@dsm-mail.saclay.cea.fr}
\affiliation{Service de Physique Th\'eorique, DSM, CE Saclay, F-91191 
Gif/Yvette, France}

\author{K. Kat${\bar o}$}
  \email{kato@nucl.sci.hokudai.ac.jp}
\author{A. Ohnishi}
  \email{ohnishi@nucl.sci.hokudai.ac.jp}
\affiliation{Division of Physics, Graduate School of Science, Hokkaido 
University, 
Sapporo 060-0810, Japan}

\author{S.M.A. Rombouts}
  \email{Stefan.Rombouts@rug.ac.be}
\affiliation{Ghent University, UGENT Department of Subatomic and Radiation 
Physics, Proeftuinstraat 86, B-9000 Gent, Belgium}

\date{\today}

\begin{abstract}

We show how every bound state of a finite system of identical fermions, 
whether a ground state or an excited one, defines a density functional. 
Degeneracies created by a symmetry group can be trivially lifted by a 
pseudo-Zeeman effect. When complex scaling can be used to regularize a 
resonance into a square integrable state, a DF also exists.

\end{abstract}


\maketitle

The aim of density functional (DF) theory is to contruct a functional
that provides the energy expectation value for a correlated many-body state
as a function of the one-body density, such that minimization of the
DF leads to the exact ground state (GS) density. Since the existence theorem 
proven for GSs by Hohenberg and Kohn (HK) \cite{HK}, its extension by Mermin 
\cite{Mer} to equilibrium at finite temperatures, and the further development 
by Kohn and Sham (KS) \cite{KS} of an equivalent, effective, independent
particle problem, a considerable amount of work has been dedicated to
generalizations such as spin DFs \cite{GunLun}, functionals taking into
account the symmetries of the Hamiltonian \cite{Goer1}, calculations of
excited state densities \cite{Goer2,LeNa}, treatments of degeneracies or
symmetries of excited states \cite{Goer3, NaLe} and quasiparticles
\cite{Neck06}. For the reader interested in an even more complete reading 
about both basic questions and applications, we refer to \cite{JD}-\cite{Gir}.

DFs for resonant states have received much less attention. We want to study 
this problem here. First we will address two related issues, namely that of a 
unified theory for ground and excited states and that of a theory for non 
degenerate and degenerate ones. A generalized existence theorem can be 
constructed by modifying the Hamiltonian in such a way that the spectrum is 
shuffled but the eigenstates are left unchanged, and by making a systematic 
use of the Legendre transform (LT) for a detailed analysis of the density.  

A reminder of the HK proof is useful here. Consider a finite number $A$ of 
identical fermions, with $a^{\dagger}_{\vec r}$ and $a_{\vec r}$ their 
creation and annihilation operators at position $\vec r,$ and the physical 
Hamiltonian, ${\bf H} = {\bf T} + {\bf V} + {\bf U}$, where 
${\bf T}=\sum_{i=1}^A t_i,$ ${\bf V}=\sum_{i>j=1}^A v_{ij}$ and 
${\bf U}=\sum_{i=1}^A u_i$ are the kinetic, two-body interaction and one-body 
potential energies, respectively. For simplicity, we consider such fermions 
as spinless and isospinless and work at zero temperature. Both $v$ and $u$ 
may be either local or non local. Next, embed the system into an additional 
one-body, external field, ${\bf W}=\sum_{i=1}^A w_i$, to observe its 
(non linear!) response. The Hamiltonian becomes ${\bf K}={\bf H}+{\bf W}$. 
It is understood that $w$ is local, 
$\langle \vec r | w | \vec r^{\, \prime} \rangle = w(r)\, 
\delta(\vec r-\vec r^{\, \prime})$, 
although a DF theory with non-local potentials exists \cite{Gil}. The usual 
Rayleigh-Ritz variational principle, where $| \psi \rangle$ is just
an $A$-particle, antisymmetric, square normalized, otherwise unrestricted wave
function, applied to $F_M=\min_{\psi}\, F,$ with $
F = \langle \psi | {\bf K} | \psi \rangle,$ generates $\psi_{min},$ the exact 
GS of ${\bf K},$ with the exact eigenvalue $F_M.$ The minimum is assumed to 
be non degenerate, smooth, reached. Clearly, $\psi_{min}$ and $F_M$ are 
parametrized by $w.$ An infinitesimal variation $\delta w$ triggers an 
infinitesimal displacement $\delta \psi_{min}$, with 
$\delta F_M = \langle \psi_{min} | \delta {\bf W } | \psi_{min} \rangle$. 
There is no first order contribution from $\delta \psi_{min}$. Define the 
one-body density matrix in coordinate representation, 
$n(\vec r,\vec r\,') = \langle \psi_{min} | a^{\dagger}_{\vec r}\, 
a_{\vec r\,'} | \psi_{min} \rangle$. Its diagonal, 
$\rho(\vec r)=n(\vec r,\vec r)$, is the usual density deduced from 
$|\psi_{min}|^2$ by integrating out all particles but one. Since 
$\delta F_M = \int d\vec r\, \rho(\vec r)\, \delta w(\vec r),$ then
$\delta F_M / \delta w(\vec r)=\rho(\vec r).$ Freeze $t,$ $v$ and $u$ and 
consider $F_M$ as a functional of $w$ alone. The HK process then consists in 
a {\it Legendre transform} of $F_M$, based upon this essential result, 
$\delta F_M/\delta w = \rho$. This LT involves two steps: i) 
subtract from $F_M$ the functional product of $w$ and $\delta F_M/\delta w$, 
{\em i.e.} the integral $\int d\vec r\, w(\vec r)\, \rho(\vec r)$, leaving 
${\cal F}_M = \langle \psi_{min} | {\bf H} | \psi_{min} \rangle$; then ii) 
set $\rho$, the ``conjugate variable of $w$'', as the primary variable rather 
than $w$; hence see ${\cal F}_M$ as a functional of $\rho$. Step ii) is 
made possible by the one-to-one ($1 \leftrightarrow 1$) map between $w$ and
$\rho$, under precautions 
such as the exclusion of trivial variations $\delta w$ that modify $w$ by a 
constant only, see for instance \cite{Leeu} and \cite{Gir}. The 
$1 \leftrightarrow 1$
map is proven by the usual argument {\it ad absurdum} \cite{HK}: if distinct 
potentials $w$ and $w'$ generated $\psi_{min}$ and $\psi_{min}^{\, \prime}$ 
(distinct!) with the same $\rho$, then two contradictory, strict inequalities 
would occur,
$\int d\vec r\, [\, w(\vec r)-w'(\vec r)\, ]\, \rho(\vec r) < 
F_M - F_M^{\, \prime},$
and,
$\int d\vec r\, [\, w(\vec r)-w'(\vec r)\, ]\, \rho(\vec r) > 
F_M - F_M^{\, \prime}.$
An inverse LT returns from ${\cal F}_M$ to $F_M$, because 
$\delta {\cal F}_M/\delta \rho=-w$. Finally, the GS eigenvalue $E_0$ of 
${\bf H}$ obtains as $E_0=\min_{\rho}\, {\cal F}_M[\rho]$; the GS wave 
function $\psi_0$ of ${\bf H}$ is the wave function $\psi_{min}$ when 
$w$ vanishes; that density providing the minimum of ${\cal F}_M$ is the 
density of $\psi_0$.

Consider now any excited bound eigenstate $\psi_n$ of ${\bf H}$, with its 
eigenvalue $E_n$. Then, trivially, $\psi_n$ is a GS of the semipositive 
definite operator $({\bf H}-E_n)^2$. Since $E_n$ is not known {\it a priori}, 
consider rather an approximate value $\widetilde E_n$, obtained by any usual 
technique (configuration mixing, generator coordinates, etc.) and assume that 
$\widetilde E_n$ is closer to $E_n$ than to any other eigenvalue $E_p$. Then 
$\psi_n$ is a GS of $({\bf H}-\widetilde E_n)^2$. The possible degeneracy 
degree of this GS is the same whether one considers ${\bf H}$ or 
$({\bf H}-\widetilde E_n)^2$. Introduce now 
$\widetilde {\bf K}=({\bf H}-\widetilde E_n)^2+{\bf W}$. If there is no 
degeneracy of either $\psi_n$ or its continuation as a functional of $w$, then 
the HK argument holds as well for $\widetilde {\bf K}$ as it does for
${\bf K}$. Hence a trivial existence proof for a DF concerning $\psi_n$.
But most often, $\psi_n$ belongs to a degenerate multiplet. Degeneracies
are almost always due to an explicitly known symmetry group of ${\bf H}$.
Notice however that the external potential $w$ does not need to show
the same symmetry; hence, in general for $\widetilde {\bf K}$, there is 
no degeneracy of its GS; a unique $\psi_{min}$ emerges to minimize the 
expectation value of $\widetilde {\bf K}$. However, for that subset of zero 
measure in the space of potentials where $w$ shows the symmetry responsible 
for the degeneracy, and in particular for the limit $w \rightarrow 0$, 
precautions are necessary. Consider therefore an (or several) additional 
label(s) $g$ sorting out the members $\psi_{ng}$ of the multiplet 
corresponding to that eigenvalue $(E_n-\widetilde E_n)^2$ of 
$({\bf H}-\widetilde E_n)^2$. There is always an operator ${\bf G}$ related 
to the symmetry group, or a chain of operators ${\bf G}_j$ in the reduction 
of the group by a chain of subgroups, which {\it commute with} ${\bf H}$ and 
can be chosen to define $g$. For simplicity, assume that one needs to consider 
one ${\bf G}$ only. Then define $g$ as an eigenvalue of ${\bf G}$ and assume, 
obviously, that the spectrum of ${\bf G}$ is not degenerate, to avoid a 
reduction chain of subgroups. It is obvious that, given some positive constant 
$C$, and given any chosen $\gamma$ among the values of $g$, there is no 
degeneracy for the GS of $({\bf H}-\widetilde E_n)^2+C\, ({\bf G}-\gamma)^2$. 
Nor is there a degeneracy of the GS of 
$ \overline {\bf K}=({\bf H}-\widetilde E_n)^2+C\, ({\bf G}-\gamma)^2+{\bf W}=
\widetilde {\bf K}+C\, ({\bf G}-\gamma)^2, $ even if $w$ has the symmetry. 
When several labels become necessary with a subgroup chain reduction, 
it is trivial to use a sum $\sum_j C_j\, ({\bf G}_j-\gamma_j)^2$ of ``pusher'' 
terms. Finally a DF results, now from the HK argument with $\overline {\bf K}$.
We stress here that pusher terms, because they commute with ${\bf H}$, 
do not change the {\it eigenstates} of either ${\bf H}$ nor 
$({\bf H}-\widetilde E_n)^2$. Only their {\it eigenvalues} are sorted out and 
reorganized. Note that the pusher expectation value vanishes for 
$\psi_{n\gamma}$. Naturally, when $w$ is finite, eigenstates of 
$\overline {\bf K}$ differ from those of $\widetilde {\bf K}$, but what 
counts is the information given by the DF when $w$ vanishes.
A simplification, avoiding cumbersome square operators ${\bf H}^2$, is worth 
noticing. Consider the operator, 
$\widehat {\bf K}={\bf H}+C\, ({\bf G}-\gamma)^2+{\bf W}.$ At the limit where 
$w$ vanishes, there is always a choice of a positive constant $C$ which makes 
the {\it lowest} state with quantum number $\gamma$ become the GS. This leads 
to a more restricted density functional that is of interest for the study of 
an yrast line.

That DF, ${\cal F}_M [\rho]$, based upon $\overline {\bf K}$, provides the 
expectation value, ${\cal F}_M [\rho] = 
\langle \psi_{min} |\, [\, ({\bf H}-\widetilde E_n)^2
             + C \, ({\bf G}-\gamma)^2\, ]\, | \psi_{min} \rangle,$
where $\psi_{min}$, square normalized to unity, is also constrained by the 
facts that 
$\langle \psi_{min} | a^{\dagger}_{\vec r} a_{\vec r} | \psi_{min} \rangle = 
\rho(\vec r)$ 
and 
$\overline {\bf K} | \psi_{min} \rangle = \varepsilon | \psi_{min} \rangle$
for the eigenvalue $\varepsilon = F_M$. 
It may be interesting to find a DF that provides the expectation value 
of ${\bf H}$ itself. This can be done by taking the 
derivative of ${\cal F}_M [\rho]$ with respect to $\widetilde E_n$, 
{\it at constant} $\rho$. We suppose that this derivative exists, which is
the case for a discrete spectrum at least. With the notation 
$|\dot{\psi}\rangle = \mbox{d} |\psi\rangle/\mbox{d} \widetilde E_n$, 
and using the fact that 
$ \langle \psi_{min} | {\bf W} | \dot{\psi}_{min} \rangle +
   \langle  \dot{\psi}_{min} | {\bf W} |\psi_{min} \rangle $ $=$ 
  $  \int w({\vec r}) \left(\mbox{d} \rho({\vec r})
                /{\mbox{d} \widetilde E_n} \right) d{\vec r} $ $= 0$,
one can write:
\begin{eqnarray}
 &\frac{\mbox{d} {\cal F}_M [\rho]}{\mbox{d} \widetilde E_n} 
   =   2\, \langle \psi_{min} |\left(\widetilde E_n - {\bf H}\right)| 
 \psi_{min} \rangle\, + & 
    \nonumber \\  &
     \langle \dot{\psi}_{min} | (\varepsilon-{\bf W}) | \psi_{min} \rangle
   + \langle \psi_{min} | (\varepsilon-{\bf W}) | \dot{\psi}_{min} \rangle
  &    \nonumber \\ 
  & \hspace{1cm} =   
  2\, \langle \psi_{min}|\left(\widetilde E_n - {\bf H}\right)|\psi_{min} 
 \rangle \, . &
\end{eqnarray}
Therefore we can define a new DF,
\begin{equation}
 {\cal F}_D [\rho] =  \widetilde E_n - 
         \frac{\mbox{d} {\cal F}_M [\rho]}{2 \, \mbox{d} \widetilde E_n} \, ,
\end{equation}
such that 
$ {\cal F}_D [\rho] = \langle \psi_{min} | {\bf H} | \psi_{min} \rangle$
and ${\cal F}_D [\rho_{n\gamma}] = E_n$ for the density $\rho_{n\gamma}$ of
the eigenstate $\psi_{n\gamma}$ of ${\bf H}$ at energy $E_n$.
Furthermore one finds that
$ \frac{\delta {\cal F}_D}{\delta \rho} [\rho_{n\gamma}] 
    = 0, $
because 
$ \frac{\delta \langle \psi_{min} |}{\delta \rho} {\bf H} | 
 \psi_{min} \rangle
 + \langle \psi_{min} | {\bf H} \frac{\delta |\psi_{min} \rangle}{\delta \rho}
   =  E_n   \frac{\delta \langle \psi_{min} | \psi_{min} \rangle }
 {\delta \rho} =0 $ for ${\psi_{min}=\psi_{n\gamma}}$. 
Hence the functional ${\cal F}_D [\rho]$ is stationary at the exact density 
$\rho=\rho_{n\gamma}$. It is not expected to be minimal at $\rho_{n\gamma}$, 
however, unless the resulting eigenstate corresponds to the absolute GS when 
$w$ vanishes.

Resonances may be defined as special eigenstates of ${\bf H}$ if one uses an 
argument {\it \`a la Gamow}, allowing some radial Jacobi coordinate 
$r \ge 0$ to show a diverging, exponential increase of the resonance wave 
function at infinity of the form $\exp(ipr)$, where the channel momentum $p$
is complex and $\Im p <0$. It is well known that those eigenvalues $E_n$ 
describing resonances are complex numbers, with $\Im E_n < 0$. There have 
been extensive discussions in the literature about the physical, or lack of, 
meaning of such non normalizable wave functions and about the wave packets 
which might be used to replace them, \cite{KF,FGR,ML,Berg}. The point of view 
we adopt in this note is based upon the Complex Scaling Method (CSM) 
\cite{Ho,Moys,Kato,GKO}: a modest modification of ${\bf H}$ transforms narrow 
resonances into {\it square integrable} states; then there is no difference 
between the diagonalization for bound states and that for resonances. The cost
of the CSM, however, is a loss of hermiticity: the CSM Hamiltonian 
${\bf H}'$ is non hermitian, somewhat similar to an optical Hamiltonian 
\cite{Ho,Moys,Kato,GKO}.

Given the ket eigenstate equation, $({\bf H}' - E_n) | \psi_n \rangle=0$,
where $| \psi_n \rangle$ is now a square integrable resonance wave function, 
we can consider the hermitian conjugate equation, 
$\langle \psi_n | ({\bf H}'^{\dagger} - E_n^*) =0$. Clearly, $\psi_n$ is a GS,
as both a ket and a bra, of the hermitian and semipositive definite operator, 
${\bf Q}_{exact}= ({\bf H}'^{\dagger} - E_n^*)\, ({\bf H}' - E_n)$, with 
eigenvalue $0$. Applying the same argument as before, but now to 
${\bf Q}_{exact}$ instead of $({\bf H}-\widetilde E_n)^2$, demonstrates the 
existence of a DF around the targeted resonant state.

In practice we do not know $E_n$ exactly. Given a sufficiently close estimate 
$\widetilde E_n$ of $E_n$, an approximate GS eigenvalue 
$| E_n-\widetilde E_n |^2$ occurs for 
${\bf Q}_{apprx}=({\bf H}'^{\dagger} - \widetilde E_n^*)\, 
                 ({\bf H}' - \widetilde           E_n  )$, 
at first order with respect to 
$\Delta {\bf Q}={\bf Q}_{apprx}-{\bf Q}_{exact}$. 
Since $\psi_n$ is not a ket eigenstate of 
${\bf H}'^{\dagger}={\bf H}'- 2 i\, \Im {\bf H'}$, 
it is also perturbed at first order in $\Delta {\bf Q}$. 
Still one can copy the construction for ${\cal F}_D [\rho],$ see Section 3,
if one interprets the operator $\mbox{d}/\mbox{d}\widetilde E_n^*$ as 
$\mbox{d}/\mbox{d}\Re \widetilde E_n + i\mbox{d}/\mbox{d}\Im \widetilde E_n$.
The resulting functional ${\cal F}_D [\rho]$ is linear in ${\bf H}'$.
For $\widetilde E_n=E_n$ the functional will be stationary at 
the density of the exact resonant state. While provinding a proof of 
existence, the construction of the exact functional for ${\bf H}'$ requires 
the knowledge of the exact eigenvalue $E_n$. This might be an inconvenient 
limitation but fortunately calculations of numbers such as $E_n$ are usually 
much easier and much more precise than calculations of wave functions 
$\psi_n$ and/or of their densities.

If the resonance has good quantum numbers (QN)s inducing degeneracies, the
same pusher terms as those which have been discussed above can be added to
create a unique GS, from the operator,
${\bf Q}_{exact} + C\, ({\bf G}-\gamma)^2$.
The HK argument, implemented with the full operator,
$\overline {\bf K}'= ({\bf H}'^{\dagger} -  E_n^*)\, 
                     ({\bf H}'           -  E_n  ) + 
C\, ({\bf G}-\gamma)^2 + {\bf W},$
then proves that DFs exist for those resonances regularized by the CSM. 
Notice, however, that a simplified theory, with an ``yrast suited'' operator 
$\widehat {\bf K}',$ linear with respect to ${\bf H}'$, is not available 
here, since the restoration of hermiticity forces a product 
${\bf H}'^{\dagger}\, {\bf H}'$ upon our formalism.

We now consider a special case of rather wide interest in nuclear and atomic 
physics. i) Good parity of eigenstates of 
${\bf H}_0={\bf T}+{\bf V}$ or ${\bf H}={\bf H}_0+{\bf U}$ when $u$ is 
restricted to be even, is assumed in the following. Hence our eigendensities,
quadratic with respect to the states, have positive parities. 
ii) We also assume that the number of fermions is even. 
iii) The QNs in which we are interested in this Section are 
the integer angular momentum $L$ and magnetic label $M$ of an eigenstate 
$\Psi_{LM}$ of ${\bf H}$, where it is understood that the two-body $v$ and 
one-body $u$ interactions conserve angular momentum. 
When $w$ is switched on and is not rotationally invariant, eigenstates of 
${\bf K}$, $\widetilde {\bf K}$, or $\overline {\bf K}$ may still tolerate 
such labels $LM$ by continuity.
First, consider $w=0$. The density $\rho_{LM}$ comes from the product 
$\Psi_{LM}^{\ *} \Psi_{LM}$, but it does not transform under rotations as an 
$\{LM\}$ tensor. Rather, it is convenient to define ``auxiliary densities'',
$\sigma_{\lambda 0}(\vec r) = \sum_{M=-L}^L (-)^{L-M} 
\langle L\ -M\ L\ M\ |\ \lambda\ 0 \rangle\ \, \rho_{LM}(\vec r)$,
where $\langle L\ -M\ L\ M\ |\ \lambda\ 0 \rangle$ is a usual 
Clebsch-Gordan coefficient. 
Each function $\sigma_{\lambda 0}(\vec r)$ now behaves under rotations 
as a $\{\lambda 0\}$ tensor. 
It can therefore be written as the product of a spherical harmonic and a
radial form factor,
$\sigma_{\lambda 0}(\vec r) = Y_{\lambda 0}(\hat r)\ \tau_{\lambda}(r) = 
\sqrt{(2\lambda+1)/4 \pi}\, {\cal L}_{\lambda}(\cos \beta)\ \tau_{\lambda}(r)
$,
where ${\cal L}_{\lambda}$ is a Legendre polynomial and the angle $\beta$ is 
the usual polar angle, counted from the $z$-axis. 
Conversely,
\begin{equation}
\rho_{LM}(\vec r) =  \sum_{\lambda=0}^{2L} (-)^{L-M}
 \langle L -M L M | \lambda 0 \rangle\,  Y_{\lambda 0}(\hat r) \,
\tau_{\lambda}(r).
\label{Fourier}
\end{equation}
This provides a ``Fourier analysis'' of $\rho_{LM}$ in angular space. 
The density is parametrized  by {\it scalar} form factors, $\tau_{\lambda}$. 
Since $L$ is here an integer and furthermore $\rho_{L-M}$ and $\rho_{LM}$ are
equal, and since Clebsch-Gordan coefficients have the symmetry property
$\langle L \ M \ L'\ M'\ |\ \lambda\ M'' \rangle\  = (-)^{L+L'-\lambda}
\, \langle L'\ M'\ L \ M \ |\ \lambda\ M'' \rangle$, 
then necessarily $\tau_{\lambda}=0$ if $\lambda$ is odd. 
There are thus $(L+1)$ scalar functions, $\tau_0$, $\tau_2$, ..., $\tau_{2L}$,
to parametrize $(L+1)$  distinct densities 
$\rho_{L0}$, $\rho_{L1}$, ..., $\rho_{LL}$. 
Because of the quadratic nature of the density observable, the even label 
$\lambda$ for angular ``modulation'' of $\rho$ runs from zero to {\it twice} 
$L$, with a ``$2L$ cut-off''; a signature, necessary if not sufficient, for an
``$L$-density''. Reinstate now $w$ as the LT conjugate of $\rho_{LM}$. 
It makes sense to study situations where $w$ is restricted to expansions 
with $(L+1)$ arbitrary scalar form factors, 
$w(\vec r)=\sum_{even\, \lambda=0}^{2L} Y_{\lambda 0}(\hat r)\,
w_{\lambda}(r).$ With inessential factors such as 
$(-)^{L-M} \langle L\ -M\ L\ M\ |\ \lambda\ 0 \rangle$ omitted for simplicity
in the following, every pair $\{ r\, \tau_{\lambda}, r\, w_{\lambda} \}$ is 
conjugate. An eigendensity of ${\bf K},\widetilde {\bf K},\overline {\bf K}$
may have an infinite number of multipole form factors, but, with such 
restricted potentials $w$, only $\tau_0, \tau_2,...,\tau_{2L}$ are chosen by 
the LT relating ${\cal F}_M$ and $F_M.$

It can make even more sense to restrict $w$ to one multipole only,
$w(\vec r)=Y_{\lambda 0}(\hat r)\, w_{\lambda}(r),$ with $\lambda=0,$ or $2,$
... or $2L,$ to study each multipole of $\rho$ separately. For simplicity we 
now use the easier version of the theory, with that operator 
$\widehat {\bf K}$ which is suited to the yrast line. Add therefore to 
${\bf H}$ a pusher term ${\bf Z}_{LM}$ leaving intact the eigenstates, 
namely
${\bf Z}_{LM}=B\, [ \vec {\bf L} \cdot \vec {\bf L} - L(L+1) ]^2 + 
C\, \left( {\bf L}_z - M \right)^2$. 
Hence $\widehat{\bf K}_{LM\lambda}={\bf H}+{\bf Z}_{LM}+{\bf W}_{\lambda}=
 {\bf T}+{\bf V}+{\bf U}+{\bf Z}_{LM}+{\bf W}_{\lambda}.$
Here the subscript $\lambda$ specifies that $w$ is reduced to one multipole 
only. Then $\vec {\bf L}$ is the total angular momentum operator and 
${\bf L}_z$ is its third component. This operator ${\bf Z}_{LM}$ moves the 
eigenvalues of ${\bf H}$ so that the lowest eigenstate of ${\bf H}$ with 
quantum numbers $\{LM\}$ becomes the GS of ${\bf H}+{\bf Z}_{LM}$. The 
commutator $[{\bf H},{\bf Z}_{LM}]$ vanishes indeed, and given $A$, $t$, $v$
and $u$, there are always positive, large enough values for $B$ and $C$ that
reshuffle the spectrum such that the lowest $\{LM\}$ eigenstate $\Psi_{LM}$ 
becomes the GS of ${\bf H}+{\bf Z}_{LM}$ under this Zeeman-like effect. 
We stress again that ${\bf Z}_{LM}$ changes nothing in the eigenfunctions, 
eigendensities, etc., of all our Hamiltonians if $w$ is rotationally 
invariant. Furthermore, angular momentum numbers remain approximately valid 
for eigenstates of $\widehat {\bf K}_{LM\lambda}$ if $w$ is weak, and the
same numbers might still make sense as labels by continuity when stronger
deformations occur. Then the usual {\it ad absurdum} argument generates a 
map $w_{\lambda} \leftrightarrow \tau_{\lambda}$, where $\tau_{\lambda}(r)$
is the form factor of the $\lambda$-multipole component of the GS density for
$\widehat {\bf K}_{LM\lambda}$, leading to an {\it exact} DF, for every
$\{LM\}$ lowest state and every even ${\lambda}$ between $0$ and $2L$. A
generalization to operators $\overline {\bf K}_{LM\lambda}$, involving
$({\bf H}-\widetilde E_n)^2$, is trivial.

This note offers theorems for the existence of exact DFs for every
excited bound state, and even narrow resonances, and every set of good
QNs used in nuclear, atomic and molecular physics. Furthermore, the
densities used as arguments of our DFs do not need to be fully
three-dimensional ones; they can be radial form factors of multipole
components of the states under study.
Our existence theorems, though, suffer from the usual plague of the field: 
constructive algorithms are missing and empirical approaches will have to be 
designed. What is the corresponding (KS) theory \cite{KS}? In its
usual form, the task of calculating the kinetic energy part of the DF is
actually left to the solution of a Schr\"odinger equation, and this can be
trivially generalized to any one-body part. Published studies of the KS
formalism are actually dedicated to calculations of the functional derivative,
$\delta {\cal V}_{xc}/\delta \rho(\vec r)$, of the exchange and correlation
part of the DF, coming from the {\it two-body} part ${\cal V}$ of the DF.
Our present use of modified Hamiltonians, or even squares of ${\bf H}$, 
introduces two-body operators, but also three- and four-body operators. 
For the versions where no squares of ${\bf H}$ occur, see the yrast suited
operator $\widehat {\bf K}$ and Section 3, the nature of the three- and
four-body terms, typically coming from 
$(\vec {\bf L}.\vec {\bf L})^2$, is not forbidding, because of
obvious factorization properties. Hence a KS theory might be realizable for
such simplified versions. With squared Hamiltonians, however, a KS theory
seems out of reach at present. A systematic analysis of solvable models on a
basis of ``modes'' \cite{Gir}, however, may help to extrapolate such models
into practical rules. For the discussion of differentiability,
representability and fine topological properties of the $w$- and 
$\rho$-spaces, we refer again to \cite{Leeu}. Up to our understanding of the
topology of the variational spaces, flat or curved \cite{Courb}, of
general use in nuclear, atomic and molecular theory, the validity domain of
our existence theorems is quite large. We have not used the time dependent
formalism, although much progress has been made in deriving excitation
energies from it \cite{PGG}. A generalization of our arguments to finite
temperatures seems plausible, however, and insofar as inverse temperature may
be viewed as an imaginary time, a generalization to a time dependent theory
is not excluded.

B.G.G. thanks T. Duguet and J. Meyer for stimulating discussions, and also
the hospitality of the Hokkaido University and Kyoto Yukawa Institute
for part of this work. S.M.A.R. thanks D. Van Neck for interesting discussions.

\end{document}